\documentclass{article}
\usepackage{amsmath,amssymb,amsfonts}
\usepackage{graphicx}

\def\const{\mathrm{const}}

\newtheorem{theorem}{Theorem}

\newtheorem{corollary}{Corollary}
\newtheorem{lemma}{Lemma}

\newtheorem{example}{Example}

\newenvironment{proof}[1][Proof]{\noindent\textit{#1.} }{\hfill$\Box$\newline\medskip}

\title{On Closed Geodesics on Ellipsoids}

\date{}

\begin{document}

\maketitle

\centerline{\large Vladimir Dragovi\'c and Milena
Radnovi\'c\footnote{on leave at the Weizmann Institute of Science,
Rehovot, Israel}}

\medskip

\centerline{Mathematical Institute SANU}

\centerline{Kneza Mihaila 35, 11000 Belgrade, Serbia and
Montenegro}

\smallskip

\centerline{e-mail: {\tt vladad@mi.sanu.ac.yu,
milena@mi.sanu.ac.yu}}

\

\begin{abstract}
Closed geodesic lines on an ellipsoid in $d$-dimensional Euclidean
space are considered. Explicit algebro-geometric condition for
closedness of such a geodesic is given. The obtained condition is
discussed in light of theta-functions theory and compared with
some recent related results.
\end{abstract}

\section{Introduction}

Geodesic motion on the ellipsoid is one of the most celebrated and
most important classical integrable systems. It has remained in
the focus interest of many researchers for almost two centuries.

\smallskip

Introducing elliptical coordinates, Jacobi proved that the
geodesic motion on the ellipsoid is integrable \cite{Jac}.
Weierstrass explicitly integrated Jacobi problem for
two-dimensional ellipsoid in terms of theta-functions
\cite{Weier}.

\smallskip

Twenty-century mathematicians also gave a great contribution to
the knowledge on geodesic motion on ellipsoid. Let us mention
results of Kn\"orrer \cite{Kn2, Kn} and Moser \cite{Mo}, where the
deep connection between Jacobi and Neumann problems is descovered
and interpreted in the modern language of isospectral theory. In
\cite{Kn}, the explicit integration of Jacobi problem in arbitrary
dimension is given.

\smallskip

Let us mention the connection between billiard motion inside an
ellipsoid and Jacobi problem and the contribution of the authors
in describing periodic billiard trajectories in
\cite{DR1,DR2,DR3,DR4}.

\smallskip

In this paper, we give analytical description of closed geodesic
lines on an ellipsoid in $d$-dimensional Euclidean space. The
article is organized as follows: in Section \ref{sec:kvadrike}, we
list the necessary prerequisites on geometry of quadrics and their
geodesic lines, in Section \ref{sec:uslovi} the analytical
conditions for closed geodesics are derived, and in Section
\ref{sec:teta} the relations between our conditions and the
theta-functions theory are discussed.

\section{Confocal Quadrics and the Caustics of a Geodesic Line on the Ellipsoid in $\mathbf R^{\mathbf
d}$}\label{sec:kvadrike}

Consider an ellipsoid in ${\mathbb R}^d$:
$$
\frac {x_1^2}{a_1}+\dots + \frac {x_d^2}{a_d}=1,\quad
a_1>\dotsb>a_d>0,
$$
and the related system of Jacobian elliptic coordinates
$(\lambda_1,\dots, \lambda_d)$ ordered by the condition
$$
\lambda_1>\lambda_2>\dotsb> \lambda_d.
$$
If we denote:
$$
Q_{\lambda}(x)=\frac {x_1^2}{a_1-\lambda}+\dots + \frac
{x_d^2}{a_d-\lambda},
$$
then any quadric from the corresponding confocal family is given
by the equation of the form:
\begin{equation}\label{eq:konfokalna.familija}
\mathcal Q_{\lambda}:\ Q_{\lambda}(x) = 1.
\end{equation}

The famous Chasles theorem states that any line in the space
$\mathbb R^d$ is tangent to exactly $d-1$ quadrics from a given
confocal family (see for example \cite{Ar}). Next lemma gives an
important condition on these quadrics.

\begin{lemma}\label{lema:polinom}
Suppose a line $\ell$ is tangent to quadrics
 $\mathcal Q_0,\mathcal Q_{\alpha_1},\dots,\mathcal Q_{\alpha_{d-2}}$
from the family {\rm (\ref{eq:konfokalna.familija})}. Then
Jacobian coordinates $(\lambda_1,\dots, \lambda_d)$ of any point
on $\ell$ satisfy the inequalities $\mathcal P(\lambda_s)\ge 0$,
$s=1,\dots,d$, where
$$
\mathcal
P(x)=-x(a_1-x)\dots(a_d-x)(\alpha_1-x)\dots(\alpha_{d-2}-x).
$$
\end{lemma}

\begin{proof} Let $x$ be a point of $\ell$,
$(\lambda_1,\dots,\lambda_d)$ its Jacobian coordinates, and $y$ a
vector parallel to $\ell$. The equation $Q_{\lambda}(x+ty)=1$ is
quadratic with respect to $t$. Its discriminant is:
$$
\Phi_{\lambda}(x,y) =
Q_{\lambda}(x,y)^2-Q_{\lambda}(y)\bigl(Q_{\lambda}(x)-1\bigr),
$$
where
$$
Q_{\lambda}(x,y) = \frac{x_1y_1}{a_1-\lambda}+\dots +
\frac{x_dy_d}{a_d-\lambda}.
$$
By \cite{Mo},
$$
\Phi_{\lambda}(x,y)=\frac
{(\alpha_1-\lambda)\dots(\alpha_{d-1}-\lambda)}
{(a_1-\lambda)\dots(a_d-\lambda)}.
$$
For each of the coordinates $\lambda=\lambda_s$, ($1\le s\le d$),
the quadratic equation has a solution $t=0$; thus, the
corresponding discriminants are non-negative. This is obviously
equivalent to $\mathcal P(\lambda_s)\ge0$.
\end{proof}

It is well known that, for a given geodesic on $\mathcal Q_0$, all
its tangent lines touch, besides $\mathcal Q_0$, the same $d-2$
quadrics $\mathcal Q_{\alpha_1}$, \dots, $\mathcal
Q_{\alpha_{d-2}}$ from the confocal family
(\ref{eq:konfokalna.familija}), see \cite{Ar, Jac}. We shall refer
to these quadrics as caustics of the geodesic line.

\smallskip

The caustics cut out several domains on the ellipsoid $\mathcal
Q_0$. Due to the Lemma \ref{lema:polinom}, the corresponding
geodesic line can be placed only in some of the domains.

\smallskip

Denote by $\Omega$ a domain on $\mathcal Q_0$, such that its
boundary $\partial\Omega$ lies in the union of confocal quadrics
$\mathcal Q_{\alpha_1}$, \dots, $\mathcal Q_{\alpha_{d-2}}$ from
the family (\ref{eq:konfokalna.familija}), and that there is a
geodesic line in $\Omega$ with caustics $\mathcal Q_{\alpha_1}$,
\dots, $\mathcal Q_{\alpha_{d-2}}$.

\smallskip

For any fixed geodesic line $\mathbf g$ on $\mathcal Q_0$ with the
caustics $\mathcal Q_{\alpha_1}$, \dots, $\mathcal
Q_{\alpha_{d-2}}$, and for any $s=1,\dots, d-1$, denote by
$\Lambda_s(\mathbf g)$ the set of all values taken by the
coordinate $\lambda_s$ on the geodesic line and
$$
\Lambda_s'=\{\,\lambda\in[a_{s+1},a_s]\, :\, \mathcal
P(\lambda)\ge0\,\}.
$$

According to Lemma \ref{lema:polinom}, we have the following:

\begin{corollary}
$\Lambda_s(\mathbf g)\subset\Lambda_s'$.
\end{corollary}

The converse is also true.

\begin{lemma}\label{lema:Lambda}
For a given geodesic line $\mathbf g$,
$$
\Lambda_s(\mathbf g)\supset\Lambda_s'
$$
for any $s=1,\dots,d-1$.
\end{lemma}

\begin{proof}
By \cite{Kn}, each of the intervals $(a_{s+1}, a_s)$, $(2\le s\le
d-1)$ contains at most two of the values
$\alpha_1,\dots,\alpha_{d-2}$, while none of them is included in
$(-\infty,a_d)\cup(a_1,+\infty)$. Thus, for each $s$, the
following three cases are possible:

\smallskip

\noindent{\it First case:} $\alpha_i,\alpha_j\in[a_{s+1},a_s]$,
$\alpha_i<\alpha_j$. Since any line tangent to the geodesic line
touches $\mathcal Q_{\alpha_i}$ and $\mathcal Q_{\alpha_j}$, the
whole geodesic is placed between these two quadrics. The elliptic
coordinate $\lambda_s$ has critical values at points where the
geodesic touches one them, and remains monotonous elsewhere.
Hence, meeting points with $\mathcal Q_{\alpha_i}$ and $\mathcal
Q_{\alpha_j}$ are placed alternately along the geodesic and
$\Lambda_s=\Lambda_s'=[\alpha_i,\alpha_j]$.

\smallskip

\noindent{\it Second case:} Among $\alpha_1,\dots,\alpha_{d-2}$,
only $\alpha_i$ is in $[a_{s+1},a_s]$. $\mathcal P$ is
non-negative in exactly one of the intervals: $[a_{s+1},
\alpha_i]$, $[\alpha_i,a_s]$, let us take in the first one. Then
the coordinate $\lambda_s$ has critical values at meeting points
with the hyperplane $x_{s+1}=0$ and the caustic $\mathcal
Q_{\alpha_i}$, and remains monotonous elsewhere. Hence,
$\Lambda_s=\Lambda_s'=[a_{s+1},\alpha_i]$. If $\mathcal P$ is
non-negative in $[\alpha_i,a_s]$, then we obtain
$\Lambda_s=\Lambda_s'=[\alpha_i,a_s]$.

\smallskip

\noindent{\it Third case:} The segment $[a_{s+1},a_s]$ does not
contain any of values $\alpha_1$, \dots, $\alpha_{d-2}$. Then
$\mathcal P$ is non-negative in $[a_{s+1},a_s]$. The coordinate
$\lambda_s$ has critical values only at meeting points with the
hyperplanes $x_{s+1}=0$, $x_s=0$ and changes mo\-no\-to\-nously
between them. This implies that the geodesic line meets them
alternately. Obviously, $\Lambda_s=\Lambda_s'=[a_{s+1},a_s]$.
\end{proof}

Denote $[\gamma_s',\gamma_s'']:=\Lambda_s=\Lambda_s'$. Notice that
the geodesic line meets quadrics of any pair $\mathcal
Q_{\gamma_s'}$, $\mathcal Q_{\gamma_s''}$ alternately. Thus, any
closed geodesic has the same number of intersection points with
each of them.

\section{Analytical Conditions for Closed Geodesic Lines on
Ellipsoid}\label{sec:uslovi}

Before formulating a criterion for sufficient and necessary
condition for closedness of real geodesic lines on the ellipsoid,
let us define the following projection of the Abel-Jacobi map.
Consider a hyperelliptic curve
\begin{equation}\label{eq:kriva}
\Gamma \ :\ y^2=\mathcal P(x),
\end{equation}
together with the standard basis of holomorphic differentials:
$$
\omega^{st}=\left[\frac{dx}y,\quad\frac{xdx}y,\quad\dots\quad,\frac{x^{d-1}dx}y\right].
$$
Denote
$$
\bar{\mathcal{A}}(P)= \left(\begin{array}{c}
0 \\
\int_0^P \dfrac{x dx}y  \\
\int_0^P \dfrac{x^2 dx}y   \\
\dots \\
\int_0^P \dfrac{x^{d-1} dx}y
\end{array}\right).
$$

\begin{theorem}\label{th:zatvorena.geodezijska}
A geodesic line on the ellipsoid $\mathcal Q_0$, with caustics
$\mathcal Q_{\alpha_1}$, \dots, $\mathcal Q_{\alpha_{d-2}}$, is
closed with exactly $n_s$ intersection points with each of
quadrics $\mathcal Q_{\gamma_s'}$, $\mathcal Q_{\gamma_s''}$,
$(1\le s\le d-1)$ if and only if
\begin{equation}\label{eq:zatvorena.geod}
\sum_{s=1}^{d-1} 2n_s\bigl(\bar{\mathcal A}(P_{\gamma_s'})
-\bar{\mathcal A}(P_{\gamma_s''})\bigr)=0.
\end{equation}
Here,
$$
 [\gamma_s',\gamma_s'']=\{\,\lambda\in[a_{s+1},a_s]\, :\, \mathcal P(\lambda)\ge0\,\}
$$
$P_{\gamma_s'}$, $P_{\gamma_s''}$ are the points on $\Gamma$ with
coordinates
 $P_{\gamma_s'}=\left(\gamma_s', (-1)^s\sqrt {\mathcal P(\gamma_s')}\right)$,
 $P_{\gamma_s''}=\left(\gamma_s'', (-1)^s\sqrt {\mathcal P(\gamma_s'')}\right)$.
\end{theorem}

\begin{proof} By \cite{Jac}, the system of differential equations
of a geodesic line on $\mathcal Q_0$ with the caustics $\mathcal
Q_{\alpha_1}$, \dots, $\mathcal Q_{\alpha_{d-2}}$ is:
\begin{equation}\label{eq:sistem}
 \sum_{s=1}^{d-1}
 \frac{\lambda_s d\lambda_s}{\sigma_s\sqrt{\mathcal P(\lambda_s)}}=0,
 \quad
 \sum_{s=1}^{d-1}\frac{\lambda_s^2 d\lambda_s}{\sigma_s\sqrt{\mathcal P(\lambda_s)}}=0,
 \quad \dots, \quad
 \sum_{s=1}^{d-1}\frac{\lambda_s^{d-1} d\lambda_s}{\sigma_s\sqrt{\mathcal P(\lambda_s)}}=0,
\end{equation}
with the same sign $\sigma_s\in\{-1,1\}$ in all of the
expressions, for any fixed $s$. Also,
\begin{equation}\label{eq:duzina}
 \sum_{s=1}^{d-1}\frac{\lambda_s^d d\lambda_s}{\sqrt{\mathcal P(\lambda_s)}}=2d\ell,
\end{equation}
where $d\ell$ is the length element.

\smallskip

Attributing all possible combinations of signs
$(\sigma_1,\dots,\sigma_{d-1})$ to $\sqrt{\mathcal P(\lambda_1)}$,
\dots, $\sqrt{\mathcal P(\lambda_{d-1})}$, we can obtain $2^{d-2}$
non-equivalent systems (\ref{eq:sistem}), which correspond to
$2^{d-2}$ different tangent lines to $\mathcal Q_0$, $\mathcal
Q_{\alpha_1}$, \dots, $\mathcal Q_{\alpha_{d-2}}$ from a generic
point of the ellipsoid $\mathcal Q_0$. Moreover, the systems
corresponding to a line and its reflection to a given
hyper-surface $\lambda_s=\const$ differ from each other only in
signs of the roots $\sqrt{\mathcal P(\lambda_s)}$.

\smallskip

Solving (\ref{eq:sistem}) and (\ref{eq:duzina}) as a system of
linear equations with respect to
$\dfrac{d\lambda_s}{\sqrt{\mathcal P(\lambda_s)}}$, we obtain:
$$
\dfrac{d\lambda_s}{\sqrt{\mathcal P(\lambda_s)}}=
\frac{2d\ell}{\prod_{i\neq s} (\lambda_s-\lambda_i)}.
$$
Thus, along the geodesic line, the differentials
$(-1)^{s-1}\dfrac{d\lambda_s}{\sqrt{\mathcal P(\lambda_s)}}$ stay
always positive, if we assume that the signs of the square roots
are chosen appropriately.

\smallskip

From these remarks and the discussion preceding this theorem, it
follows that the value of the integral
$\int\dfrac{\lambda_s^id\lambda_s}{\sqrt{\mathcal P(\lambda_s)}}$
between two consecutive common points of the geodesic and the
quadric $\mathcal Q_{\gamma_s'}$ (or $\mathcal Q_{\gamma_s''}$) is
equal to:
$$
2(-1)^{s-1} \int_{\gamma_s'}^{\gamma_s''}
\dfrac{\lambda_s^id\lambda_s}{+\sqrt{\mathcal P(\lambda_s)}}.
$$

Now, if $\mathbf g$ is a closed geodesic having exactly $n_s$
points at $\mathcal Q_{\gamma_s'}$ and $n_s$ at $\mathcal
Q_{\gamma_s''}\ (1\le s\le d-1)$, then
$$
\sum \int^{\mathbf g} \dfrac{\lambda_s^id\lambda_s}{\sqrt{\mathcal
P(\lambda_s)}}= 2\sum(-1)^{s-1}n_s
\int_{\gamma_s'}^{\gamma_s''}\dfrac{\lambda_s^id\lambda_s}{+\sqrt{\mathcal
P(\lambda_s)}}, \quad (1\le i\le d-2).
$$
Finally, the geodesic line is closed if and only if
$$
\sum (-1)^s
2n_s\int_{\gamma_s'}^{\gamma_s''}\dfrac{\lambda_s^id\lambda_s}{\sqrt{\mathcal
P(\lambda_s)}}=0, \quad (1\le i\le d-2),
$$
which was needed.
\end{proof}

\begin{example}
{\rm An interesting class of closed geodesic lines on the
ellipsoid in three-dimensional space, in obtained in \cite{Fe1}.}
\end{example}

\begin{corollary}
If a geodesic line on the ellipsoid $\mathcal Q_0$, with caustics
$\mathcal Q_{\alpha_1}$, \dots, $\mathcal Q_{\alpha_{d-2}}$,
satisfies the condition
\begin{equation}\label{eq:zatvorena.dovoljan}
  \sum_{s=1}^{d-1} n_s \bigl(\mathcal A (P_{\gamma_s'})-\mathcal
  A(P_{\gamma_s''})\bigr)=0,
\end{equation}
then it is closed with exactly $n_s$ intersection points with each
of quadrics $\mathcal Q_{\gamma_s'}$, $\mathcal Q_{\gamma_s''}$,
$(1\le s\le d-1)$.
\end{corollary}

\begin{proof}
After the reparametrization:
$$
ds=\lambda_1\lambda_2\cdots\lambda_dd\ell,
$$
the equation (\ref{eq:duzina}) transforms into:
\begin{equation}\label{eq:duzina2}
\sum_{s=1}^{d-1} \dfrac{d\lambda_s}{\sqrt{\mathcal
P}(\lambda_s)}=2ds.
\end{equation}
Now, this equation, together with the system (\ref{eq:sistem}) is
equivalent to the condition (\ref{eq:zatvorena.dovoljan}).
\end{proof}

Let us note that the condition (\ref{eq:zatvorena.dovoljan}) will
be satisfied if and only if (\ref{eq:zatvorena.geod}) holds (i.e.\
the geodesic line is closed) and its length $L$ with respect to
the parameter $s$:
$$
ds=\lambda_1\lambda_2\cdots\lambda_nd\ell
$$
is such that the vector
$$
\left(
\begin{array}{c}
L/2\\
0\\
\vdots\\
0
\end{array}\right)
+\sum_{s=1}^{d-1} n_s\bigl(\bar{\mathcal A}(P_{\gamma_s'})
-\bar{\mathcal A}(P_{\gamma_s''})\bigr)
$$
belongs to the period-lattice of the Jacobian of the corresponding
hyper-elliptic curve.

\section{Closed Geodesics and Theta Functions}\label{sec:teta}

In this section, we are going to present a different approach to
finding an analytical condition for the closed geodesics on the
ellipsoid, based on results of \cite{Dubr}.

\smallskip

The equations of a geodesic line on the ellipsoid in the
$d$-dimensional space are:
\begin{equation}\label{eq:geodezijska}
\aligned
 x_1(t)&=\alpha_1\frac{\theta[\alpha_1,\beta_1](tU+z_0)\theta(z_0)}{\theta[\alpha_1,\beta_1](z_0)\theta(tU+z_0)},\cr
 x_2(t)&=\alpha_2\frac{\theta[\alpha_2,\beta_2](tU+z_0)\theta(z_0)}{\theta[\alpha_2,\beta_2](z_0)\theta(tU+z_0)},\cr
 &\vdots\cr
 x_d(t)&=\alpha_d\frac{\theta[\alpha_d,\beta_d](tU+z_0)\theta(z_0)}{\theta[\alpha_d,\beta_d](z_0)\theta(tU+z_0)}.
\endaligned
\end{equation}
Here,
$$
\alpha_i=\bigl[\prod_{j\neq i}(a_i-a_j)\bigr]^{-1/2},
$$
the theta-functions are constructed over the Riemann surface
(\ref{eq:kriva}), $z_0$ is an arbitrary vector of the Jacobian
$\mathcal J(\Gamma)$, and $U$ is a vector of $b$-periods of the
differential $\Omega$ of the second order with the pole at the
point $x=\infty$:
$$
 \Omega = \frac{x^{d-1}+a_{d-1}z^{d-1}+\dots+a_0}{2\sqrt{\mathcal P(x)}}dx,
$$
normalized by the condition
$$
 \oint_{a_i}\Omega=0,\quad
 U_i=\oint_{b_i}\Omega\quad (i=1,\dots, d-1).
$$
Pairs $(\alpha_i, \beta_i)$ are the corresponding characteristics
(for more detailed explanation, see \cite{Dubr} and \cite{Mum}).

\smallskip

A sufficient condition for the periodicity of the curve given by
(\ref{eq:geodezijska}) is:
$$
TU=n_1E_1+n_2E_2+\dots n_gE_g+m_1F_1+m_2F_2+\dots m_gF_g,\quad
g=d-1
$$
for some $T>0$, where $E_1$,... $E_g$, $F_1$,..., $F_g$ are the
basis vectors of the period lattice of a basis $\omega$ of
holomorphic differentials, and $n_k$, $m_j$ are integers.

\smallskip

We can calculate explicitly the vector $U$ in coordinate system
associated with some basis $\omega$ of holomorphic differentials.
The order of the pole of $\Omega$ at $\infty$ is equal to $2$,
thus \cite{Dubr}:
$$
U_i=\oint_{b_i}\Omega=f_i(\infty).
$$
Here, $\omega_i=f_i(t)dt$, where $t$ is a local coordinate around
$\infty$ and $\omega=[\omega_1,\dots ,\omega_g]$.

\smallskip

For the standard basis
$$
\omega_1^{st}=\frac{dx}y,\dots,\omega_g^{st}=\frac{x^{g-1}dx}y.
$$
since $x=1/t^2$, we have around $\infty$:
$$
 \omega_k^{st}
 =\frac{-2t^{2g-2k}dt}{\sqrt{\overline{\mathcal P}(t^2)}}.
$$
Here
$$
\overline{\mathcal P}(\xi)
=-(a_1\xi-1)\dots(a_d\xi-1)(\alpha_1\xi-1)\dots(\alpha_{d-2}\xi-1).
$$
Thus, $\omega_k^{st}$ has a zero of order $2g-2k$ at the infinity
point, for $1\le k\le g-1$, and $\omega_g^{st}$ has no zero at
this point.

\smallskip

Finally, we get the formula for the vector $U$:
$$
U^{st}=[0,\dots,0, u_g]^T.
$$

\smallskip

Denote now
$$
\bar{\mathcal{A'}}(P)= \left(\begin{array}{c}
\int_0^P \dfrac{dx}y\\
\int_0^P \dfrac{x dx}y  \\
\int_0^P \dfrac{x^2 dx}y   \\
\dots \\
\int_0^P \dfrac{x^{d-2} dx}y \\
0
\end{array}\right).
$$

\begin{theorem}\label{th:d-kriterijum}
Let a geodesic line $\mathbf g$ on the ellipsoid $\mathcal Q_0$
associated with a curve (\ref{eq:kriva}) where  the ordered zeroes
of the polynomial $\mathcal P$ are
$$
z_1<\dots <z_{2g}
$$
be given. A sufficient condition for $\mathbf g$
 to be closed is
$$
\sum_{s=1}^{g} 2n_s\bigl(\bar{\mathcal A'}(z_{2s}) -\bar{\mathcal
A'}(z_{2s-1})\bigr)+\sum_{s=1}^{g} m_s\bigl(\bar{\mathcal
A'}(b_{s})\bigr) =0,
$$
where $b_s$ denotes a basis of $b$-cycles and basis of $a$-cycles
of $\Gamma$ is represented by double cuts $[z_{2s-1},z_{2s}]$.
\end{theorem}

\subsection*{Acknowledgments}
The research was partially supported by the Serbian Ministry of
Science and Technology, Project {\it Geometry and Topology of
Manifolds and Integrable Dynamical Systems}. The authors would
like to thank Prof.\ B.\ Dubrovin, Yu.\ Fedorov and S.\ Abenda for
interesting discussions. One of the authors (M.R.) acknowledges
her gratitude to Prof.\ V.\ Rom-Kedar and the Weizmann Institute
of Science for the kind hospitality and support during in the work
on this paper.

\end{document}